\documentclass[%
 preprint
 amsmath,amssymb,
 aps, physrev,
]{revtex4-2}

\usepackage{graphicx}
\usepackage{dcolumn}
\usepackage{bm}
\usepackage{subcaption}
\usepackage{amsmath}
\usepackage{bm}

\begin{document}

\preprint{APS/123-QED}

\title{\textbf{On the decay instability of electron acoustic waves} 
}%

\author{F. Valentini}
 \email{Contact author: francesco.valentini@unical.it}
\affiliation{%
 Dipartimento di Fisica, Universit\'a della Calabria, Rende (CS), Italia
}%
\author{T. M. O'Neil and D. H. Dubin }
\affiliation{%
 Physics Department, University of California at San Diego, La Jolla (CA), USA
}%

\date{\today}

\begin{abstract}
Electron acoustic waves (EAWs) are nonlinear plasma modes characterized by electron trapping, which suppresses the usual Landau damping. Despite being predicted in the 1990s, their excitation and decay mechanisms remain a subject of active research. This study investigates the nonlinear dynamics of EAWs, focusing on their excitation, decay instability, and the role of vortex merging in phase space. Using Eulerian Vlasov-Poisson simulations, we reproduce the excitation of stable EAWs via an external resonant driving force and explore their decay under the effect of a low-amplitude perturbation. The study identifies a distinct $2 \to 1$ decay process, where an EAW with a shorter wavelength merges into a longer-wavelength mode, driven by vortex dynamics in phase space. We find that the instability is triggered by a small fraction of particles capable of transitioning between potential wells, facilitating energy exchange between two adjacent phase space holes and vortex merging. Our simulations highlight the chaotic nature of particle trajectories in the vicinity of the separatrix between trapped and free phase space regions, which significantly contributes to the instability growth. Additionally, we analyze the influence of the perturbation amplitude on the growth rate of the instability, shedding light on the critical role of phase-space dynamics in the decay process. These findings offer a deeper understanding of the nonlinear behavior of plasma waves and suggest future directions for studying plasma wave stability in more complex systems, as the decay mechanism discussed here is likely to be universal in plasmas with closed separatrices in phase space, underscoring its significance in nonlinear plasma dynamics.
\end{abstract}

\maketitle

\section{Intrduction}

In 1991, Holloway and Dorning \cite{holloway1991} observed that certain nonlinear wave structures could manifest within a plasma even when the amplitude is low. They termed these structures "electron acoustic waves" (EAWs) due to the acoustic-like form of their dispersion relation, which is described by $\omega = 1.31 k v_{th}$ for small values of the wave number \( k \). Here, \( \omega \) represents the angular frequency of the wave, \( k \) denotes the wave number, and \( v_{th} \) is the thermal velocity of the plasma electrons. According to linear wave theory, EAWs would experience significant Landau damping, since their phase velocity is comparable to the thermal velocity of electrons \cite{landau1946}. Despite this, EAWs are nonlinear modes, characterized by electrons that become trapped in the wave troughs. This trapping effect \cite{oneil1965} results in a flattened electron velocity distribution at the wave’s phase velocity, effectively suppressing Landau damping.

The mechanism of EAW excitation was further elucidated in Ref. \cite{valentini2006}, where the authors demonstrated that these waves can be initiated by an external small-amplitude driving force, provided that the force is applied resonantly over several electron trapping periods and they are, in fact, Bernstein-Green-Kruskal (BGK) nonlinear modes \cite{bernstein1957}. Moreover, in 2009 the excitation of EAWs has been successfully obtained in experiments with nonneutral plasmas \cite{anderegg2009}. The external resonant driver continuously compensates for the energy dissipated through Landau damping, thereby sustaining the formation of a trapped particle distribution and facilitating the emergence of the EAW. This phenomenon was investigated using particle-in-cell (PIC) \cite{valentini2006} and Eulerian \cite{valentini2008} simulations, which also revealed that, under the condition where the wavelength is the longest possible wavelength that can fit within the plasma system, the generated EAW is stable and maintains a nearly constant amplitude even after the driving force is turned off. Conversely, for cases where the EAW does not have the longest possible wavelength, a decay process occurs, leading to a transition towards a longer-wavelength EAW. In phase space, this decay is observed as a merger of vortex structures formed by trapped particles, a behavior analogous to an inverse cascade process \cite{berk1970,ghizzo1988,manfredi2000,depackh1962,dory1964}.

This decay instability and the associated merging of vortex structures have been extensively studied in the context of Trivelpiece-Gould waves \cite{trivelpiece1959}, both from the experimental \cite{affolter2018,affolter2019} and the theoretical/numerical \cite{dubin2018,dubin2019} point of view. Notably, Dubin \cite{dubin2018} has provided a detailed theoretical explanation of the decay mechanisms in Trivelpiece-Gould waves, involving weakly trapped particles, becoming detrapped and then retrapped in the wave potential well. Moreover, in Ref. \cite{dubin2024}, theory and experiments investigating the nonlinear beat wave decay of diocotron modes in a nonneutral plasma column are described.

In the present work, we employ Eulerian Vlasov-Poisson (VP) simulations to reproduce the excitation process of the EAWs and their decay instability, specifically focusing on the dynamics of trapped particles. To gain deeper insights into this latter phenomenon, we complement the analysis with the numerical integration of the motion equations of a large ensemble of test particles, moving in the electric field obtained through the Eulerian VP simulations, and demonstrate that only a very small fraction of resonant particles, whose dyanamics become chaotic near the separatrix between trapped and free phase space regions \cite{valentini2005}, are responsible for the occurrence of the instability. Our simulations provide crucial information on how and why vortex merging occurs, elucidating the underlying physical mechanisms responsible for the instability.
In particular, we focus on the $2\to 1$ decay process, in which an EAW with a wavelength corresponding to one half of the total length of the spatial numerical domain ($m=2$ Fourier component) decays  to a largest wavelength EAW ($m=1$ Fourier component). In this scenario, the use of a zero noise Eulerian VP numerical algorithm is crucial, as it shows that the instability occurs only when a small amplitude perturbation is added to the $m=1$ Fourier component of the electric field and the growth rate of the instability slightly depends on the $m=1$ perturbation amplitude. We provide an explanation for these evidences.

This paper is organized as follows. In Section II, we present the numerical model and simulation setup, detailing the Vlasov-Poisson equations, initial conditions, and numerical schemes used. Section III describes the numerical results, including the excitation of stable electron acoustic waves (EAWs), analysis of the decay instability, and the role of the perturbation in triggering vortex merging. Moreover, we examine the dynamics of test particles in phase space to gain insight into the mechanisms driving the instability. Finally, we conlude in Section IV, providing a discussion of the findings, including implications for plasma wave stability and connections to previous studies; we also suggest avenues for future research on nonlinear plasma wave dynamics and EAW stability.

\section{Numerical model and simulation setup}
In unmagnetized plasmas, within the kinetic theory framework, the behavior of high-frequency plasma oscillations is governed by the one-dimensional, nonlinear Vlasov-Poisson (VP) system of equations, which can be written in dimensionless form as follows:
\begin{eqnarray}
\label{vlasov}
& &\frac{\partial f}{\partial t} + v \frac{\partial f}{\partial x} - (E+E_D) \frac{\partial f}{\partial v} = 0\\
& &\frac{\partial E}{\partial x} =1- \int f \, dv,
\label{poisson}
\end{eqnarray}
where $f=f(x,v,t)$ is the electron distribution function, (ions are treated as a stationary background providing a neutralizing positive charge), $E=E(x,t)$ is the self-consistent electric field and $E_D=E_D(x,t)$ the external driver electric field.
In Eqs. (\ref{vlasov})-(\ref{poisson}), time is normalized to the inverse electron plasma frequency $\omega_{p}$, velocities to the electron thermal speed $v_{th}$ and lengths by the Debye length $\lambda_D$. Accordingly, electric fields are scaled by $m_e\omega_{pe}v_{th}/e$,
where $e$ and $m_e$ are the electron charge and mass, respectively. From now on all quantities will be scaled to the characteristic parameters listed above.

The numerical integration of the Vlasov equation is carried out using the well-known splitting method \cite{cheng1977} within the electrostatic approximation, coupled with a finite-difference upwind scheme \cite{vanleer1976}, ensuring third-order accuracy in both phase space and time. 
The phase space domain for the simulations is defined as 
$D = [0, L] \times [-v_{max}, v_{max}]$, where $L$ is the length of the spatial numerical domain and $v_{max}=6$ indicates the limits of the velocity domain. Periodic boundary conditions are implemented in physical space, while in velocity space $f(x,|v|>v_{max},t)=0$. Poisson equation is solved numerically using a standard Fast Fourier Transform (FFT) routine.
Typically, simulations are conducted using $N_x = 1024 - 2048$ grid points in physical space and $N_v = 2001$ grid points in velocity space. The time step $\Delta t$ is chosen to ensure compliance with the Courant-Friedrichs-Lewy (CFL) condition \cite{CFL}.
In order to control numerical accuracy of the VP simulations, an energy conservation equation derived from Eqs. (\ref{vlasov})-(\ref{poisson}) is employed. The relative variations in total energy remain always consistently smaller than $10^{-7}\%$ in all simulations presented here.

As for the numerical integration of the equations of motion of the test particles under the effect of the electric field from the VP simulations, we employed (and compared) two different approaches, the so-called Boris pusher \cite{boris1970} (massively employed for Particle-in-cell simulations) and a standard third order Adams-Bashfort multi-step method \cite{butcher2008}, obtaining almost identical results.

\section{Numerical results}
Our analysis is performed in four steps, described in detail as follows.

\subsection{Excitation of a stable EAW through an external driver}
\begin{figure}[h!]
    \centering
    \includegraphics[width=\textwidth]{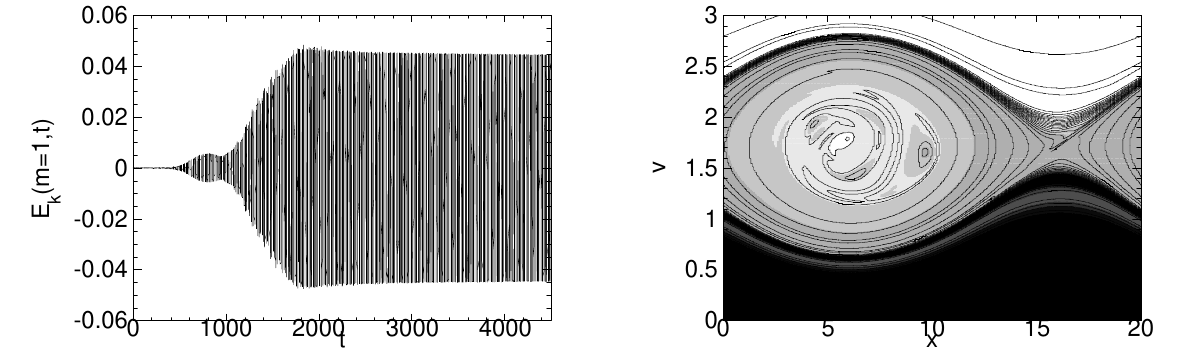}
    \caption{Time evolution of the electric field Fourier component $m=1$ (left panel)and phase space contours of the electron distribution function around the wave phase speed at $t=t_{max}$ (right panel).}
    \label{fig0}
\end{figure}

As a first step, in order to reproduce the excitation of a stable EAW, we run the exact same simulation presented in Ref. \cite{valentini2006}, in which an external driver of the form:
\begin{equation}
E_D(x,t)=E_D^{max}\left[1+\left(\frac{t-\tau}{\Delta\tau}\right)^n\right]^{-1}\sin{(kx-\omega_Dt)}
\end{equation}
where $E_D^{max}=0.01$, $\tau=1200$,  $\Delta\tau=600$, $n= 10$, $k=m(2\pi/L)$ ($m=1$, $L=20$) and $\omega_D=kv_{\varphi,D}$ ($v_{\varphi,D}=1.7$), is applied to an initial equilibrium in which electrons have homogeneous density and Maxwellian distribution of velocities. The external driver is turned on and off adiabatically, in order to avoid the excitation of Langmuir waves, and its amplitude is definitively null at $t\gtrsim 2000$; the maximum time for this simulation is $t_{max}=4500$ and $N_x=1024$.
At the end of the simulation, an EAW with the largest wavelength that fits in the spatial domain is excited and is associated with a single vortical BGK structure in phase space (see Ref.\cite{valentini2006}).
The results of this preliminary simulation are summarized in Fig. \ref{fig0}, where the time evolution of the electric field Fourier component $m=1$ is shown in the left panel, while the phase space contours of the electron distribution function around the wave phase speed are displayed in the right panel at $t=t_{max}$.

\subsection{EAW decay instability}
As a second step, we replicate this EA mode periodically in space, thus generating a distribution function with two identical phase space holes, and use it as the initial condition for a new set of simulations in a longer domain. The matching from wavelength to wavelength is smooth since periodic boundary conditions are used in the initial simulation. For these new simulations we set $L=40$, $N_x=2048$, $t_{max}=3000$ and no external driver is applied to the plasma. We performed 11 simulations, perturbing the new EAW-BGK equilibrium through a sinusoidal density disturbance of the form $A_1\cos{(kx)}$ with $k=m2\pi/L$ and low amplitude in the range $0\leq A_1\leq10^{-5}$,  imposed at $t=0$ on the Fourier component $m=1$.

\begin{figure}[h!]
    \centering
    \includegraphics[width=\textwidth]{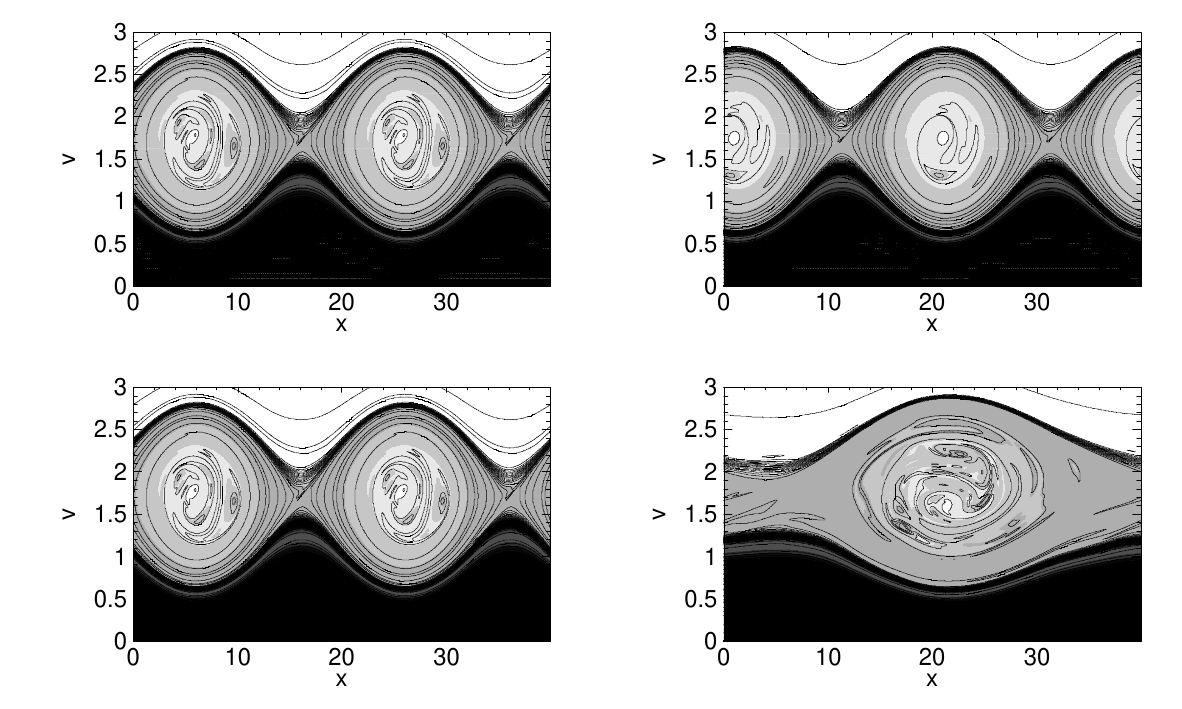}
  \caption{Phase space contours of the electron distribution function around $v_\varphi$ at $t=0$ (left column) and at $t=t_{max}$ (right column) for a simulation with $A_1=0$ (top row) and $A_1=10^{-5}$ (bottom row).}
    \label{fig1}
\end{figure}

In Fig. \ref{fig1}, we report the phase space contours of the electron distribution function around $v_\varphi$, at $t=0$ (left column) and $t=t_{max}$ (right column) for a simulation with $A_1=0$ (top row) and $A_1=10^{-5}$ (bottom row). It is evident from this figure that a vortex merging process occurs when a low amplitude perturbation is imposed on the system, while the phase space structures (and consequently the electric oscillations) are stable in absence of perturbations, on the timescale of the simulation.

\begin{figure}[h!]
    \centering
    \includegraphics[width=0.6\textwidth]{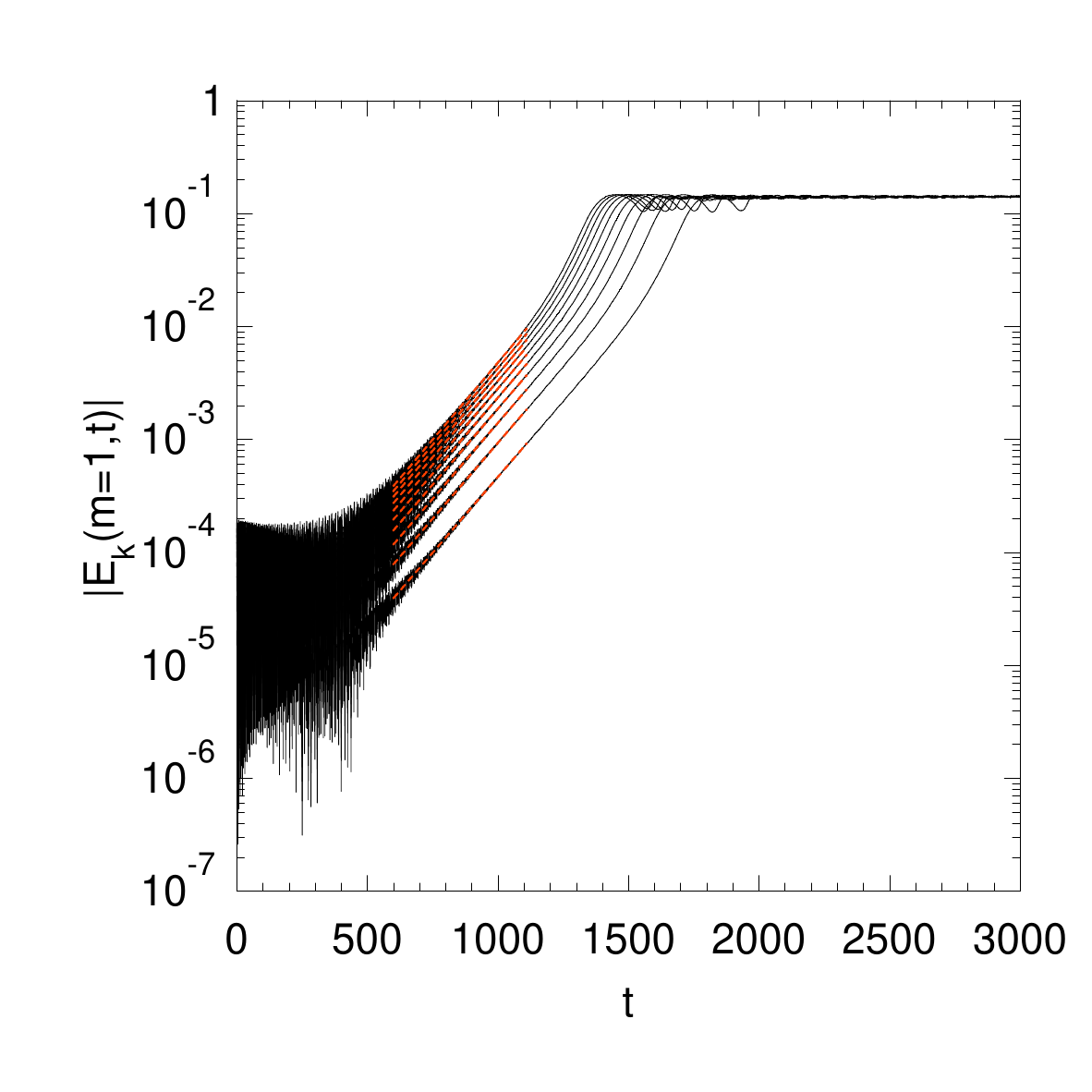}
    \caption{Time evolution of the absolute value of the $m=1$ Fourier component of the electric field on a semi-logarithmic scale across 10 simulations initialized with density perturbation and exhibiting the $2 \to 1$ instability. The red-dashed curves represent the best fits during the exponential growth phase of the instability.}
    \label{fig2}
\end{figure}

Figure \ref{fig2} displays, on a semi-logarithmic scale, the time evolution of the absolute value of the $m=1$ Fourier component of the electric field across 10 simulations, each initialized with a small amplitude perturbation on the $m=1$ Fourier mode and exhibiting the $2 \to 1$ instability. The red-dashed curves in the plot represent the best fits obtained during the exponential growth phase of the instability, from which the growth rate $\Gamma$ was estimated for each simulation. Following the growth phase, the amplitude of the fluctuations reaches a saturation level that is identical across all simulations. This uniform saturation amplitude of the electric field is attributed to the fact that the energy associated with the initial double-hole phase-space structure is the same in each simulation. Consequently, the merging process consistently results in a single vortex of uniform velocity scale, generating an electric signal with the same amplitude in each run. 

\begin{figure}[h!]
    \centering
    \includegraphics[width=0.6\textwidth]{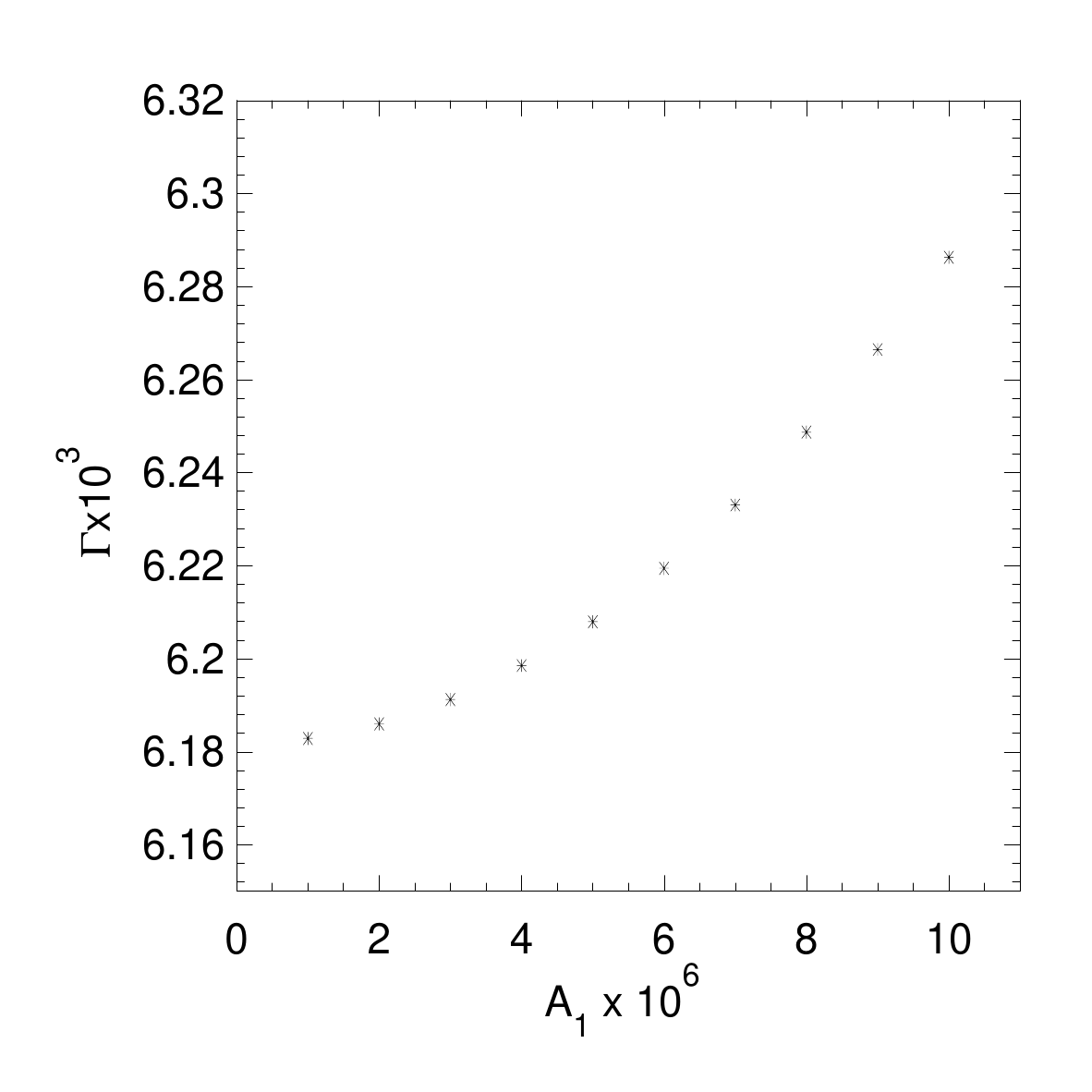}
\caption{Dependence of the growth rate $\Gamma$ on the amplitude $A_1$ of the initial density disturbance, showing a clear correlation between these two quantities.}
    \label{fig3}
\end{figure}

Figure \ref{fig3} illustrates the dependence of the growth rate $\Gamma$ on the amplitude $A_1$ of the initial density perturbation, revealing a clear correlation between these two quantities.

\subsection{Test-particles phase space dynamics}
To investigate the mechanism underlying the decay instability observed in the previous section, specifically why and how the phase-space trapping vortices tend to merge and why the growth rate of the instability slightly increases with higher initial perturbation amplitude, we numerically integrated the equations of motion for $N=60000$ test particles under the influence of the electric field from the VP simulations. At $t=0$, all particles were positioned in phase space such that they were trapped within the two phase-space holes, with an equal distribution between them.

\begin{figure}[h!]
    \centering
    \includegraphics[width=0.6\textwidth]{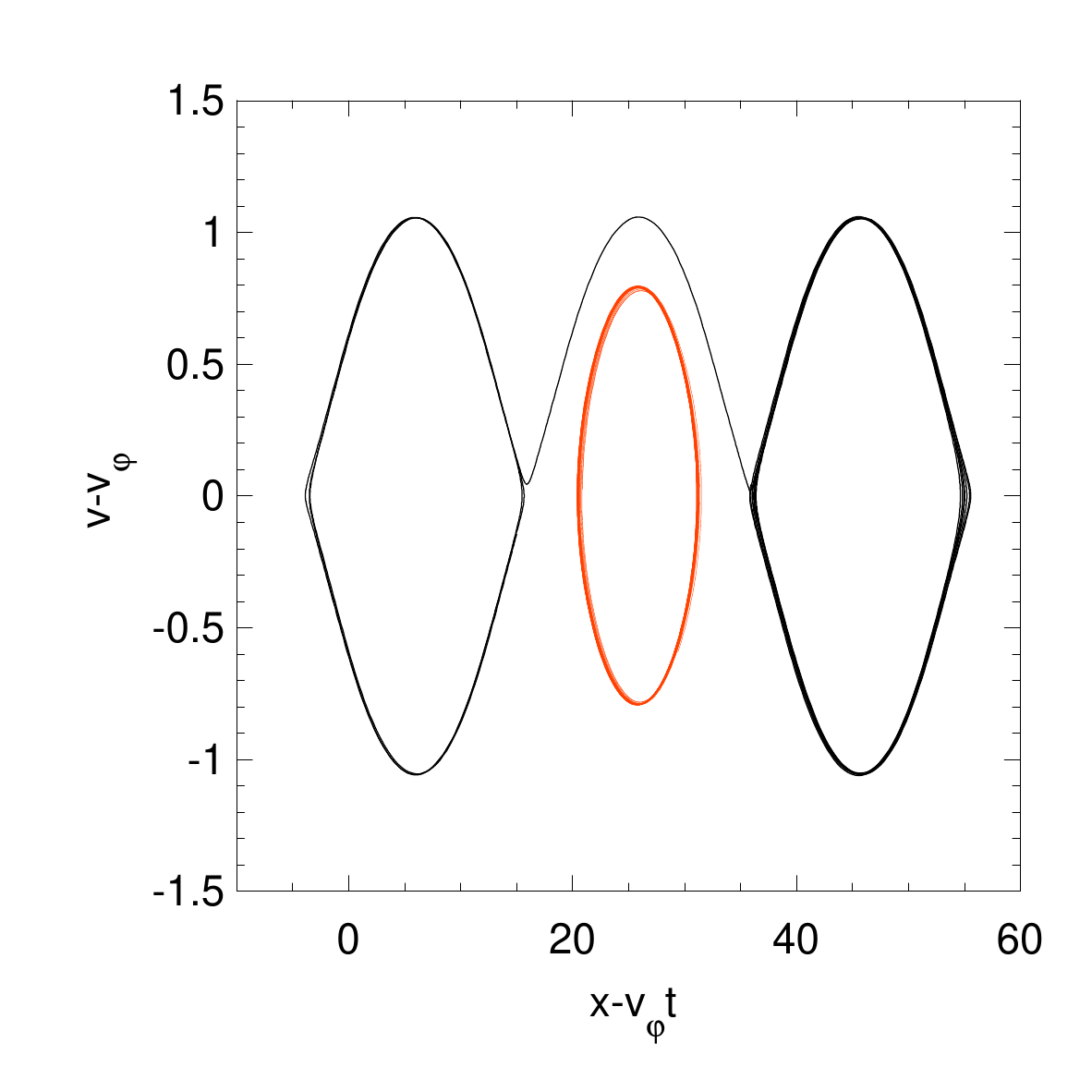}
\caption{Trajectories of two test particles from the simulation with $A_1 = 10^{-5}$ in the wave reference frame, over the time interval $0 \leq t \leq 1200$, corresponding to the exponential growth phase of the instability.}
    \label{fig4}
\end{figure}

An inspection of the phase-space trajectories revealed that, in the stable simulation (without density disturbance), all particles remained confined within the potential well where they were initially placed at $t=0$, and the two phase-space holes did not interact with each other.
The situation changes drastically when a low-amplitude density perturbation is introduced into the system at $t = 0$. Figure \ref{fig4} presents the trajectories of two test particles from the simulation with $A_1 = 10^{-5}$ in the wave reference frame, over the time interval $0 \leq t \leq 1200$, which corresponds to the exponential growth phase of the instability.
In contrast to the majority of particles that remain trapped within a single wave potential well (represented by the red curve in the plot), we observe a small fraction of particles capable of transitioning between wells (represented by the black curve), with some undertaking long excursions through phase space before being re-trapped. These particles, which jump from one phase-space hole to the other, establish a connection between the two wells and facilitate an effective energy exchange between them.

We analyze the phase-space flights of 20,000 particles over the time range $0 \leq t \leq 1200$. A phase-space flight is defined as the spatial distance traveled by a particle between two consecutive changes in the sign of its velocity, in the wave reference frame. When this distance exceeds the spatial width of each phase-space hole (half of the total length $L$), it indicates that a particle has jumped from one potential well to another. Particles remaining within a single phase-space hole exhibit phase-space flights shorter than $L/2$.

Although the number of phase-space flights with length greater than $L/2$ is small — about 0.16\% of the total flights in the case of the simulation with $A_1 = 10^{-5}$ — these flights are crucial for triggering an energy exchange between the two phase-space holes. This exchange, which occurs only when a small-amplitude density disturbance is initially imposed, creates an energetic imbalance between the holes. Consequently, one hole gains more energy (and increases in size) while absorbing particles from the other, which is ultimately swallowed.

\begin{figure}[h!]
    \centering
    \includegraphics[width=0.6\textwidth]{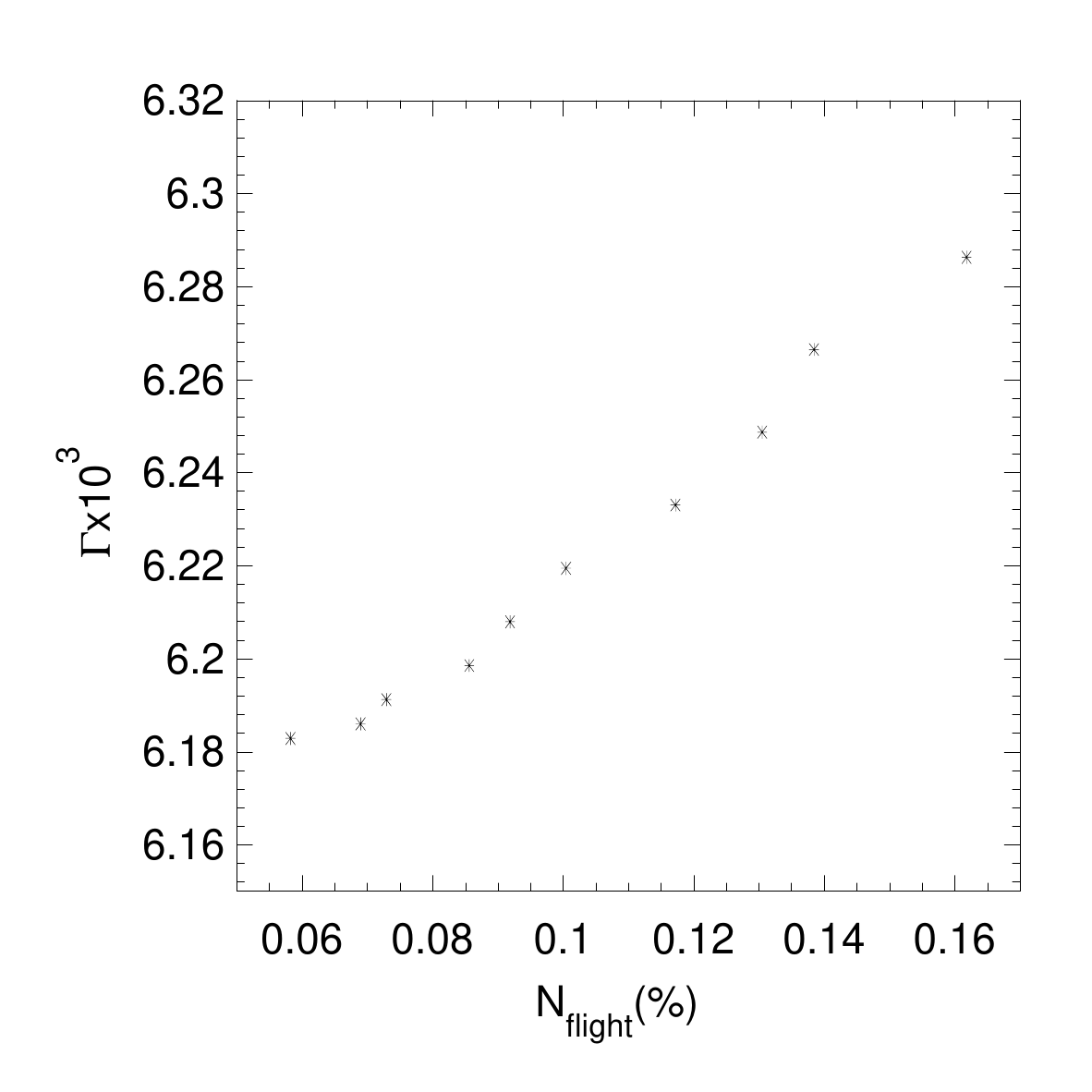}
\caption{Dependence of the growth rate $\Gamma$ on $N_{flight}$, the number of phase-space flights with length greater than $L/2$.}
    \label{fig5}
\end{figure}

Thus, the larger the number of phase-space flights with length greater than $L/2$ (i.e., the number of particles that can become detrapped and retrapped), the faster the vortex merging process and the instability. As shown in Fig. \ref{fig5}, the growth rate $\Gamma$ increases with $N_{flights}$, the number of phase-space flights with length greater than $L/2$.

Based on the considerations above, we hypothesize that the phase-space dynamics of particles capable of becoming detrapped and retrapped, and transitioning between potential wells, may exhibit chaotic behavior. To test this hypothesis, the total number of test particles was divided into three populations, each containing $20000$ particles. The initial conditions for each population were set up as follows in each simulation:

(i) Each particle in the first population is assigned an initial phase-space location $P_{0,j} = [x_{0,j}(t=0), v_{0,j}(t=0)]$, with $j = 1, \cdots, 20000$. All particles are initially trapped within the two phase-space holes, with an equal distribution between them.

(ii) Each particle in the second population is assigned an initial location $P_{1,j}$, which is separated from $P_{0,j}$ by a phase-space distance:
\begin{equation}
d\chi_j(t=0) = \sqrt{[x_{1,j}(t=0) - x_{0,j}(t=0)]^2 + [v_{1,j}(t=0) - v_{0,j}(t=0)]^2}
\end{equation}

(iii) Each particle in the third population is assigned an initial location $P_{2,j}$, separated from $P_{0,j}$ by a phase-space distance:
\begin{equation}
d\xi_j(t=0) = \sqrt{[x_{2,j}(t=0) - x_{0,j}(t=0)]^2 + [v_{2,j}(t=0) - v_{0,j}(t=0)]^2}
\end{equation}

Particles in both the second and third populations also satisfy the trapping condition.

\begin{figure}[h!]
    \centering
    \includegraphics[width=0.6\textwidth]{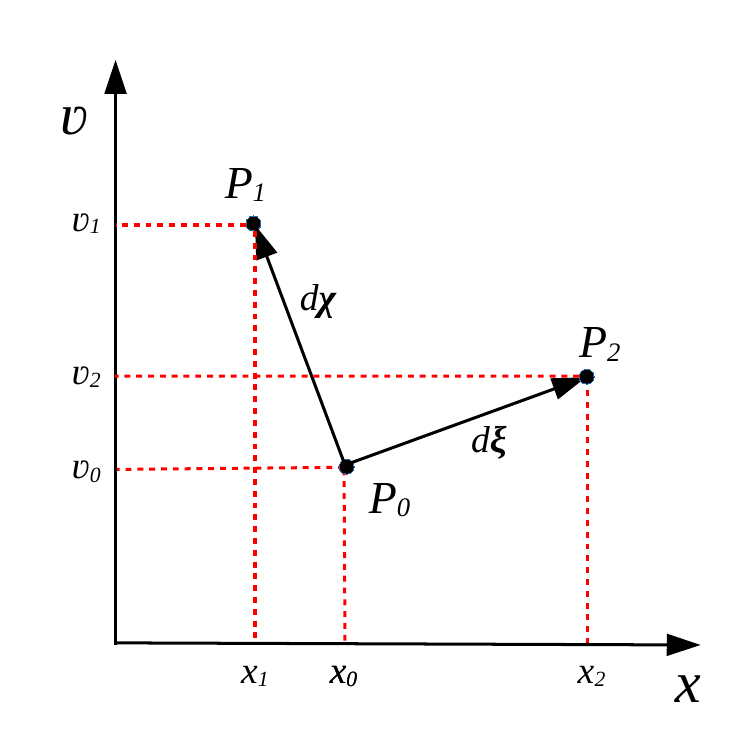}
    \caption{Schematic representation of the initial conditions \( P_0 \), \( P_1 \), and \( P_2 \) for three particles, each belonging to a different population.}
    \label{fig_star}
\end{figure}

The initial conditions \( P_0 \), \( P_1 \), and \( P_2 \) for three particles (each belonging to a different population) are schematically depicted in Fig. \ref{fig_star}. It is important to note that the two vectors \( d\bm{\chi} \) and \( d\bm{\xi} \) are initially perpendicular to each other. This choice enables us to monitor the time evolution of these distances throughout the simulation, with the objective of assessing the degree of chaos in each phase-space trajectory. The degree of chaos is associated with an exponential increase in the phase-space separation between two infinitesimally close initial conditions. Since the Vlasov equation describes an incompressible flow, the phase space volume is preserved. Therefore, if the phase-space separation increases in one direction, it must decrease in the perpendicular direction. Focusing solely on the time evolution of the separation in a single direction could overlook the direction of maximal separation. In all simulations, we set \( d\xi_j(t=0) = d\chi_j(t=0) = 10^{-6} \) for \( j = 1, \dots, 20000 \).

\begin{figure}[h!]
    \centering
    \includegraphics[width=0.6\textwidth]{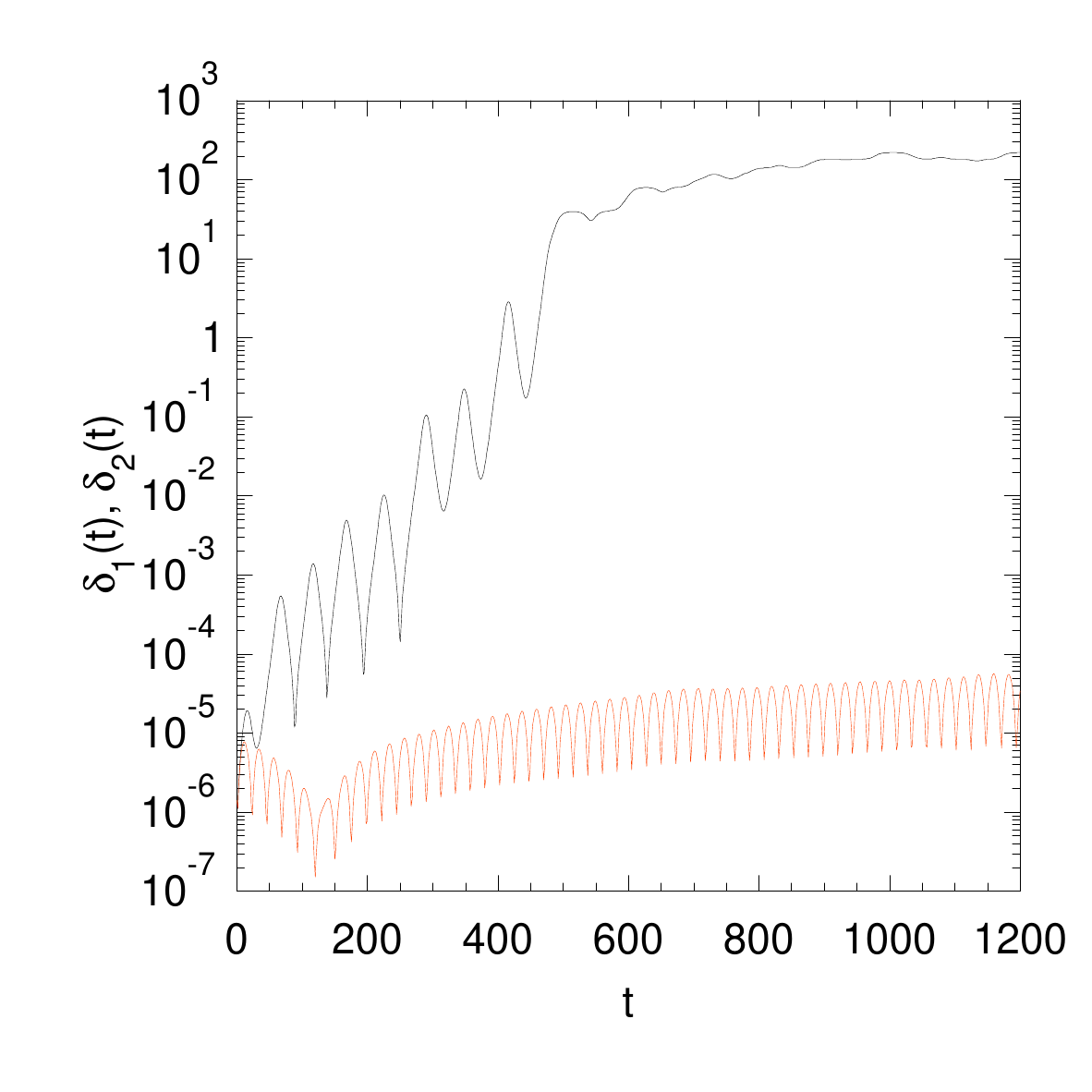}
\caption{Time dependence of \( \delta_1(t) = \max\{d\chi_1(t), d\xi_1(t)\} \) (red) and \( \delta_2(t) = \max\{d\chi_2(t), d\xi_2(t)\} \) (black) in a semi-logarithmic plot, corresponding to the trapped particle (red) and the jumping particle (black) shown in Fig. \ref{fig4}, respectively.}
    \label{fig6}
\end{figure}

Figure \ref{fig6} shows the time dependence of \( \delta_1(t) = \max\{d\chi_1(t), d\xi_1(t)\} \) (red) and \( \delta_2(t) = \max\{d\chi_2(t), d\xi_2(t)\} \) (black) in a semi-logarithmic plot, corresponding to the trapped particle (red) and the jumping particle (black) shown in Fig. \ref{fig4}, respectively. As can be seen, for the trapped particle, the phase-space separation between two infinitesimally close initial conditions remains very small throughout the simulation. In contrast, for the jumping particle, the separation increases significantly, initially displaying exponential growth in the time range \( 0 \leq t \lesssim 500 \). Beyond this time, although the growth continues, it occurs at a slower rate, indicating a deceleration in the separation evolution after \( t = 500 \).

If the \( j \)-th phase-space trajectory becomes chaotic, the corresponding phase-space separation will increase exponentially in the asymptotic time limit as:
\begin{eqnarray}
d\chi_j(t) &\simeq& d\chi_j(t=0) e^{\lambda_{1,j} t} \implies \lambda_{1,j} =\lim_{t\to\infty} \frac{1}{t} \log \left( \frac{d\chi_j(t)}{d\chi_j(t=0)} \right) \\\nonumber
\\
d\xi_j(t) &\simeq& d\xi_j(t=0) e^{\lambda_{2,j} t} \implies \lambda_{2,j} = \lim_{t\to\infty}\frac{1}{t} \log \left( \frac{d\xi_j(t)}{d\xi_j(t=0)} \right)
\end{eqnarray}
The Lyapunov exponent associated with this \( j \)-th trajectory is defined as \( \lambda_j = \max\{\lambda_{1,j}, \lambda_{2,j}\} \).

\begin{figure}[h!]
    \centering
    \includegraphics[width=0.6\textwidth]{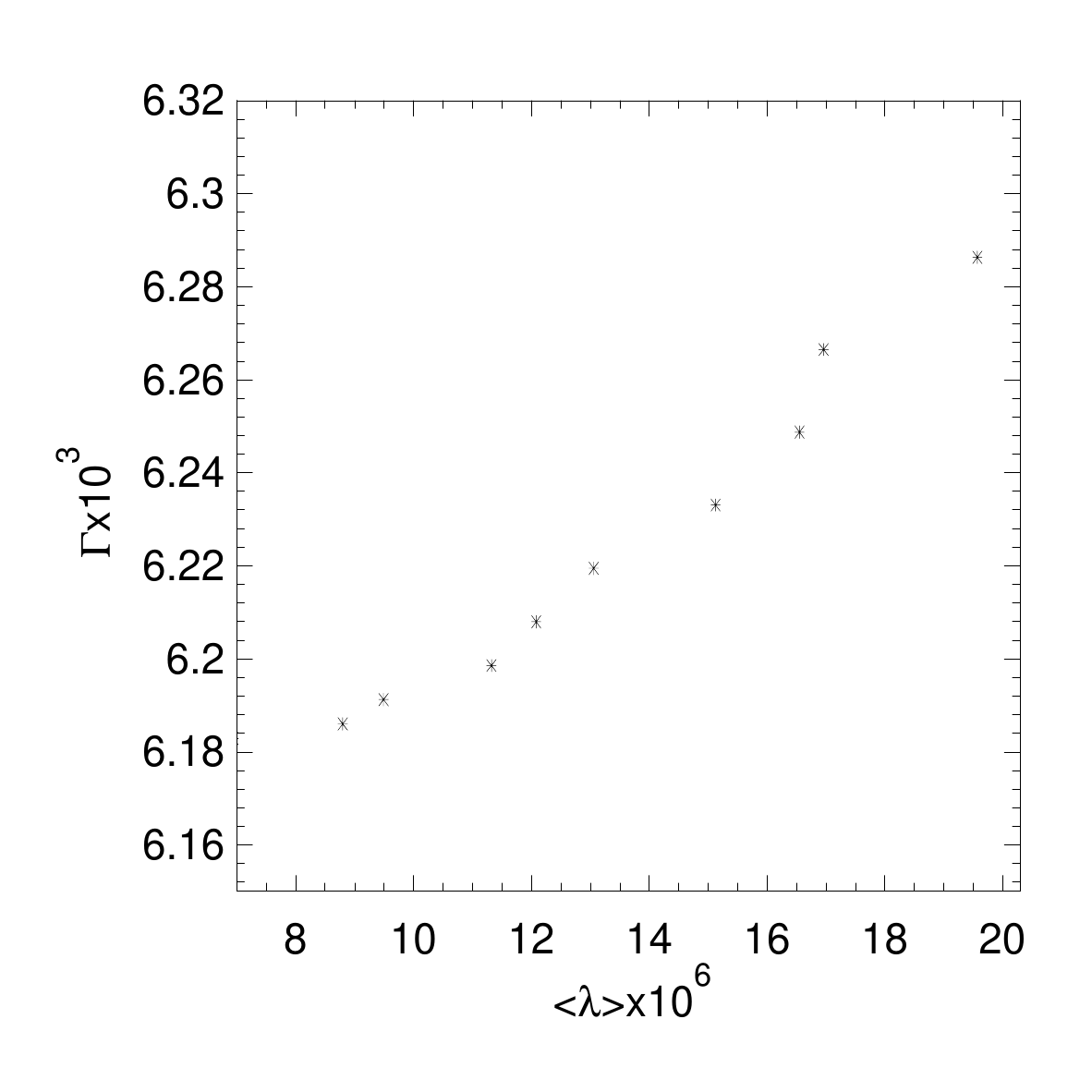}
\caption{Relationship between the growth rate \( \Gamma \) of the decay instability and the average value \(\langle \lambda \rangle\) of the Lyapunov exponents, for ten different simulations with initial perturbation amplitude in the range \(10^{-6} \leq A_1 \leq 10^{-5}\).}
    \label{fig7}
\end{figure}

For the purposes of the present work, we aim to estimate the finite-time Lyapunov exponents within the time range \(0 \leq t \leq 1200\), corresponding to the growth phase of the instability, rather than in the long-time limit. To this end, we introduce the following definitions for the \(j\)-th phase space trajectory:

\begin{eqnarray}
   & &\lambda_{1,j} = \frac{1}{N_t \Delta t} \sum_{i=0}^{N_t} \log \left( \frac{d\chi_{j,i+1}}{d\chi_{j,i}} \right) \\ \nonumber
   \\
   & &\lambda_{2,j} = \frac{1}{N_t \Delta t} \sum_{i=0}^{N_t} \log \left( \frac{d\xi_{j,i+1}}{d\xi_{j,i}} \right).
\end{eqnarray}
where \(N_t\) is the integer closest to \(1200/\Delta t\), \(d\chi_{j,i} = d\chi_j(i \Delta t)\), \(d\chi_{j,i+1} = d\chi_j((i+1) \Delta t)\), \(d\xi_{j,i} = d\xi_j(i \Delta t)\), and \(d\xi_{j,i+1} = d\xi_j((i+1) \Delta t)\). Finally, the finite-time Lyapunov exponent for the \(j\)-th trajectory is defined as \( \lambda_j = \max\{\lambda_{1,j}, \lambda_{2,j}\} \). 
By performing this calculation for each phase-space trajectory within the time interval \( 0 \leq t \leq 1200 \), we found that the Lyapunov exponents for the subset of particles capable of becoming detrapped and retrapped—those undergoing phase-space flights with spatial lengths greater than \( L/2 \)—are non-zero. This result indicates that the phase-space dynamics of these particles are indeed chaotic. Figure \ref{fig7} illustrates the relationship between the growth rate \( \Gamma \) of the decay instability and the average value \(\langle \lambda \rangle\) of the Lyapunov exponents for these particles for ten different simulations with initial perturbation amplitude in the range $10^{-6}\leq A_1\leq 10^{-5}$. Firstly, we observe that increasing \( A_1 \) leads to a higher number of particles that can become detrapped and retrapped. Consequently, this also results in an increase in \(\langle \lambda \rangle\).
Finally, the figure clearly shows that \( \Gamma \) increases as the degree of phase-space chaos, represented by \(\langle \lambda \rangle\), becomes more pronounced.

\begin{figure}[h!]
    \centering
    \includegraphics[width=\textwidth]{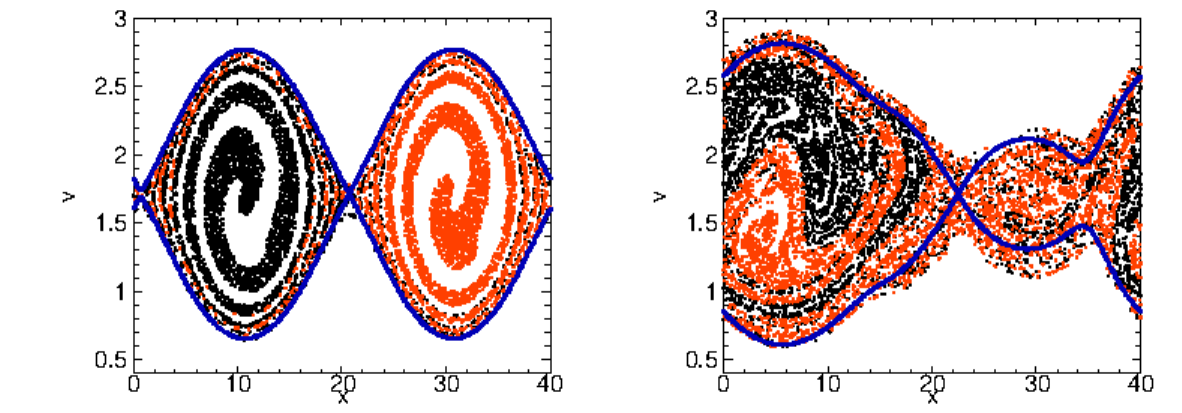}
\caption{Poincare sections at two different times in the simulation with $A_1=10^{-5}$, before (left) and during (right) the occurrence of the vortex merging process. Particles initially trapped in different potential wells have been assigned different colors (black and red dots), while the blue curves in each plot represent the separatrices.}
    \label{fig_ref}
\end{figure}

Finally, in Fig. \ref{fig_ref} we show  Poincare sections at two different times in the simulation with $A_1=10^{-5}$, before (left) and during (right) the vortex merging process takes place. We assigned different colors (red and black) to the populations of particles initially trapped in one and in the other potential well. During the time evolution, it is evident that red particles start diffusing through the separatrices (solid-blue curves) and migrate to the other well (left plot). This effect is increasingly visible as time goes on: red particles migrate to the black well and, as a consequence, the red vortex gets smaller and smaller.

\subsection{Artificially restoring the EAW stability}
\begin{figure}[h!]
    \centering
    \includegraphics[width=0.6\textwidth]{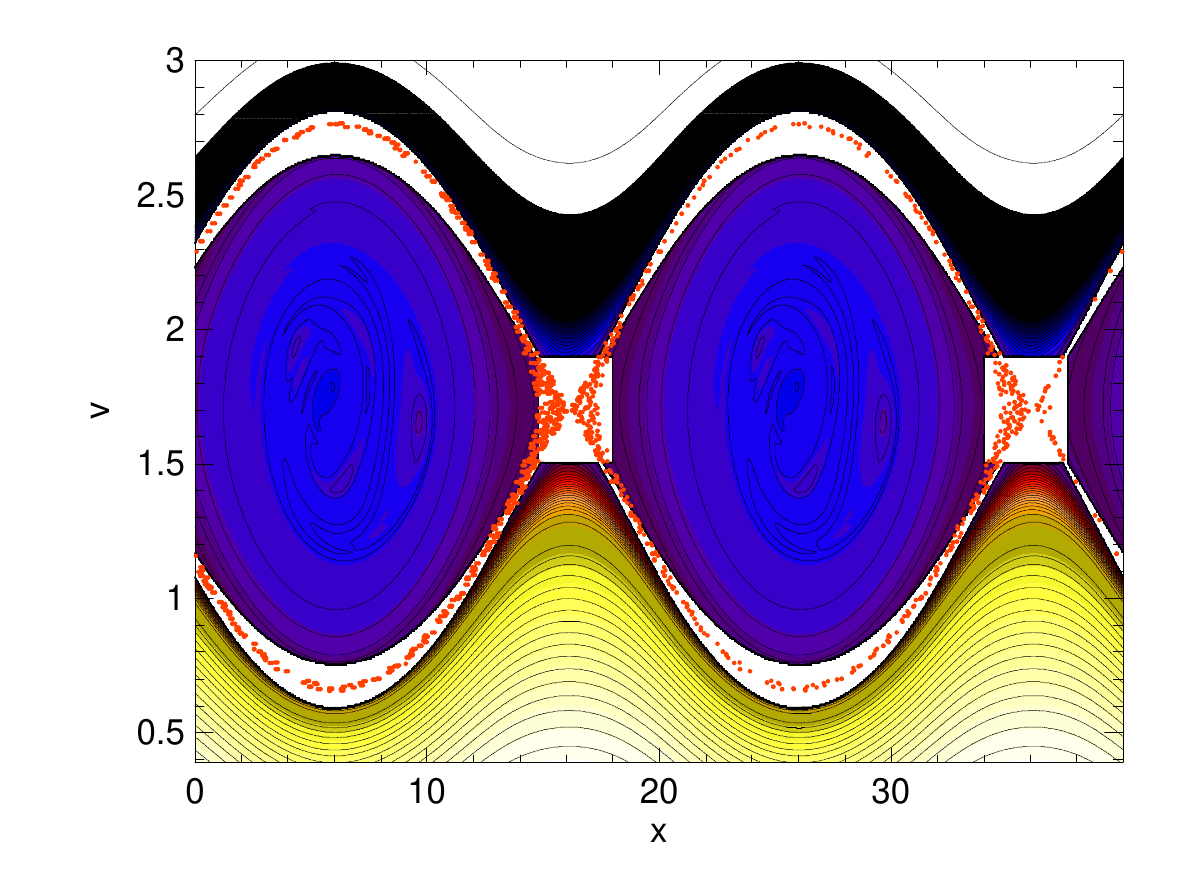}
\caption{Contour plot of the electron distribution function in phase space at \( t = 0 \), with red dots marking the initial positions of particles whose phase-space dynamics later become chaotic. The white regions, which encompass the separatrix and the \( x \)-points, represent areas where the electron distribution function has been artificially set to zero.}

    \label{fig8}
\end{figure}

At this point, it is natural to inquire where these chaotic particles were initially located at \( t = 0 \) in the simulations. We specifically examine the numerical run with \( A_1 = 10^{-5} \) and trace back the trajectories of these particles. Figure \ref{fig8} presents a contour plot of the electron distribution function in phase space at \( t = 0 \), with red dots marking the initial positions of particles whose phase-space dynamics later become chaotic. Unsurprisingly, these particles are concentrated near the separatrix between the trapped and free phase-space regions and close to the \( x \)-point between the two phase-space holes.

\begin{figure}[h!]
    \centering
    \includegraphics[width=\textwidth]{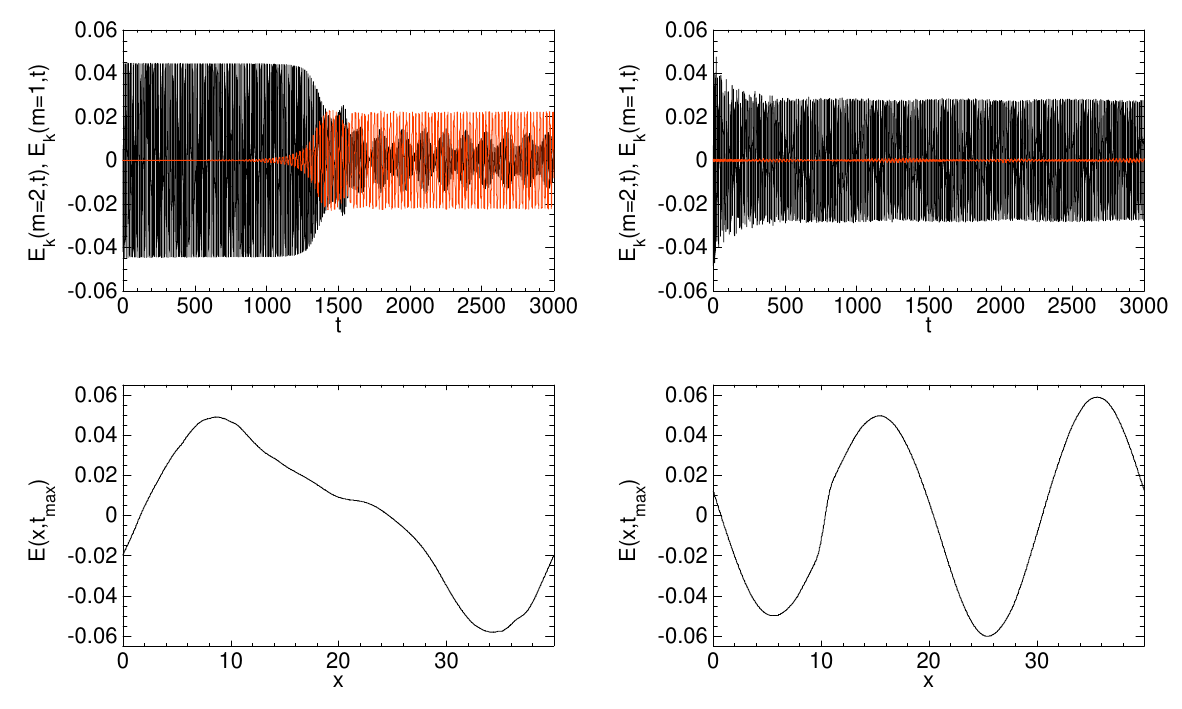}
\caption{(Top panels) Time evolution of the electric field spectral components \( m = 2 \) (black) and \( m = 1 \) (red);  (Bottom panels) electric field as a function of \( x \) at \( t = t_{\text{max}} \). Results from the unstable simulation with \( A_1 = 10^{-5} \) are reported in the left column, while those from the artificially stable simulation (also with \( A_1 = 10^{-5} \)) are displayed in the right column.}

    \label{fig9}
\end{figure}

Therefore, we artificially set the electron distribution function to zero in the phase-space regions where these particles are initially located (see in Fig. \ref{fig8} the white regions encompassing the separatrix and the \( x \)-points) and then repeat the simulation with this artificially modified initial condition. The time evolution of the electric field spectral components $m = 2$ (black) and $m = 1$ (red) is depicted in the top panels of Fig.~\ref{fig9}. In the bottom panels, the electric field as a function of $x$ at $t = t_{\text{max}}$ is shown. These results correspond to the unstable simulation with $A_1 = 10^{-5}$ (left column) and the artificially stable simulation (right column), also with $A_1 = 10^{-5}$. The latter is initialized with the artificially modified initial condition, as previously discussed, where the phase regions corresponding to particles that may become chaotic have been removed.

It is evident from this figure that, in the first case (left column), the decay instability occurs, and the mode $m=1$ grows exponentially over time. In contrast, the instability is completely suppressed in the second simulation (right column), where particles that could become detrapped and retrapped have been artificially removed. As a result, the electric field exhibits a single spatial oscillation (with the $m=1$ mode dominating) in the unstable case, while, in the artificially stable case, two spatial oscillations are observed (with the $m=2$ mode dominating).

From these observations, we can conclude that the few particles capable of transitioning from one potential well to another, whose dynamics in phase space becomes chaotic, are the primary drivers responsible for triggering the decay instability.

\section{Summary and Conclusions}

In this study, we explored the excitation and decay of electron acoustic waves (EAWs) in unmagnetized plasmas through high-resolution Eulerian Vlasov-Poisson (VP) simulations. Our investigation was motivated by prior research demonstrating that EAWs, although heavily damped in linear theory due to Landau damping, can be sustained nonlinearly by trapped particles. These trapped particles flatten the velocity distribution, minimizing Landau damping and allowing EAWs to form as Bernstein-Green-Kruskal (BGK) modes. Using an external driver applied resonantly over multiple electron trapping times, we successfully reproduced the excitation process of stable EAWs in our simulations, confirming the results presented by Valentini et al. \cite{valentini2006}. 
Our primary focus, however, was to analyze the decay instability observed in EAWs, particularly the transition from a double-hole phase space structure (characterized by the $m=2$ Fourier mode) to a single-vortex structure (characterized by the $m=1$ Fourier mode). This $2 \to 1$ decay process was systematically induced by adding a low-amplitude disturbance to the electric field at the $m=1$ mode. The results showed that the growth rate slightly increases with the initial perturbation amplitude and this decay process is highly sensitive to low-level perturbations. Phase-space analysis reveals that the decay is driven by chaotic particle dynamics near the separatrix, which allow a small fraction of resonant particles to transition between the two vortices, facilitating vortex merging. This particle dynamics, observed only when a low-amplitude perturbation is introduced, leads to an effective energy transfer between vortices, destabilizing the double-vortex configuration and resulting in a single, larger vortex.
To gain further insight into the chaotic dynamics of the particles responsible for this instability, we conducted numerical integration of test particles in the electric field generated by the VP simulations. This approach allowed us to track individual particle trajectories in phase space. We found that particles confined within a single vortex retained stable, non-chaotic trajectories, while a small fraction of particles exhibited chaotic trajectories characterized by Lyapunov exponents larger than zero. These particles undergo phase-space flights between vortices, which ultimately initiate the merging process. Our results demonstrate that the instability growth rate is directly correlated with the number of particles that execute phase-space flights, supporting our hypothesis that inter-vortex energy exchange through chaotic particle motion is the key mechanism driving vortex merging.
The detailed phase-space analysis presented here enhances our understanding of EAW decay instabilities, establishing a direct connection between initial density disturbance, chaotic particle dynamics, and vortex merging. This finding has important implications for the stability of EAWs in laboratory and space plasmas, where background noise and particle trapping can play significant roles in determining wave stability. Moreover, our results underline the critical role of chaotic dynamics near the separatrix in plasma instability, a phenomenon with potential relevance in various plasma environments.

Future work will involve extending this analysis to higher-dimensional simulations, which could capture additional degrees of freedom and provide a more comprehensive description of EAW instabilities in realistic plasma configurations. Additionally, the effect of stronger initial perturbations on the nonlinear dynamics of particle trapping and vortex merging will be explored to determine thresholds for instability suppression or enhancement in experimental setups. Finally, further investigations will be focused on including noise effects, such as weak collisions, which may increase the number of weakly trapped particles as plasma diffuses across the separatrices, presumably enhancing the growth rate of the instability. Furthermore, a systematic simulation study of growth rate dependence on the amplitude of the EAW fluctuations could provide relevant insights, refining models for amplitude-dependent growth rates observed in TG modes \cite{dubin2018} and extending them to EAWs for comparison with simulations.

The data that support the findings of this study are available from the corresponding author upon reasonable request.

\begin{acknowledgments}
This project has received funding from the European Union’s Horizon Europe research and innovation programme under grant agreement No. 101082633 (ASAP) and from the ASI project “Attività di Fase A per la missione Plasma Observatory” (2024-15-HH.0).
FV acknowledges the support of the PRIN 2022 project “The ULtimate fate of TuRbulence from space to laboratory plAsmas (ULTRA)” (2022KL38BK), funded by the Italian Ministry of University and Research.
\end{acknowledgments}

\newpage

\end{document}